\documentclass[twocolumn,preprintnumbers,amsmath,amssymb]{revtex4}
\usepackage{graphicx}
\usepackage{dcolumn}
\usepackage{bm}

\begin{document}
\title{
Current-voltage characteristics of a  graphene nanoribbon
field-effect transistor
}
\author{V.~Ryzhii\cite{byline}, 
M.~Ryzhii, and A.~Satou
}
\affiliation{
Computer Solid State Physics Laboratory, University of Aizu, 
Aizu-Wakamatsu, 965-8580, Japan}
\affiliation{
Japan Science and Technology Agency, CREST, Tokyo 107-0075, Japan
}
\author{T.~Otsuji}

\affiliation{Research Institute for Electrical Communication,
\affiliation{Tohoku University,  Sendai,  980-8577, Japan}
Japan Science and Technology Agency, CREST, Tokyo 107-0075, Japan
}
\date{\today}

\begin{abstract}
We present an analytical device model for a   field-effect
transistor based on a heterostructure which consists of
an array of nanoribbons clad between the highly conducting
substrate (the back gate) and the top gate controlling
the source-drain current. 
The equations of the model 
of  a graphene nanoribbon field-effect transistor (GNR-FET)
include the Poisson equation
in the weak nonlocality approximation.
Using this model,
we  find explicit analytical formulas
for the spatial distributions
of the electric potential along the
channel and for the  GNR-FET current-voltage characteristics 
(the dependences of the source-drain current on the drain
voltages as well as on the back gate and top gate voltages)
for  different geometric parameters of the device.
It is shown that the shortening of the top gate can result in a substantial
modification of the GNR-FET current-voltage characteristics.
\end{abstract}

\maketitle

\section{Introduction}

{\it Graphene}, i.e.,  
a monolayer of carbon atoms forming a dense honeycomb 
two-dimensional (2D) crystal structure  
is considered as a promising candidate for future micro- and nanoelectronics
~\cite{1,2,3,4,5}.
The features of the electron and hole 
energy spectra 
in graphene provide the
exceptional properties of
graphene-based heterostructures and devices,
in particular field-effect transistors~\cite{4,6}.
Fabrication and experimental studies as well as a theoretical
model of graphene field-effect transistor (G-FET)
was reported recently in, for instance, Ref.~\cite{6}.
The operation of G-FETs is accompanied by the formation
of the lateral  n-p-n  (or p-n-p) junction under the controlling (top)
gate and the pertinent energy barrier. The current through this barrier
can be associated by both thermionic and tunneling processes
~\cite{6,7,8,9,10}.
The utilization of the patterned graphene which constitutes an
array of sufficiently narrow graphene strips (nanoribbons)
provides an opportunity to engineer the band structure
to achieve the optimal device parameters.
In particular, properly choosing the width of the nanoribbons,
one can fabricate the graphene structures with relatively wide
band gap but rather high electron (hole) mobility~\cite{4,11}.
\begin{figure}
\centerline{\includegraphics[width=75mm]{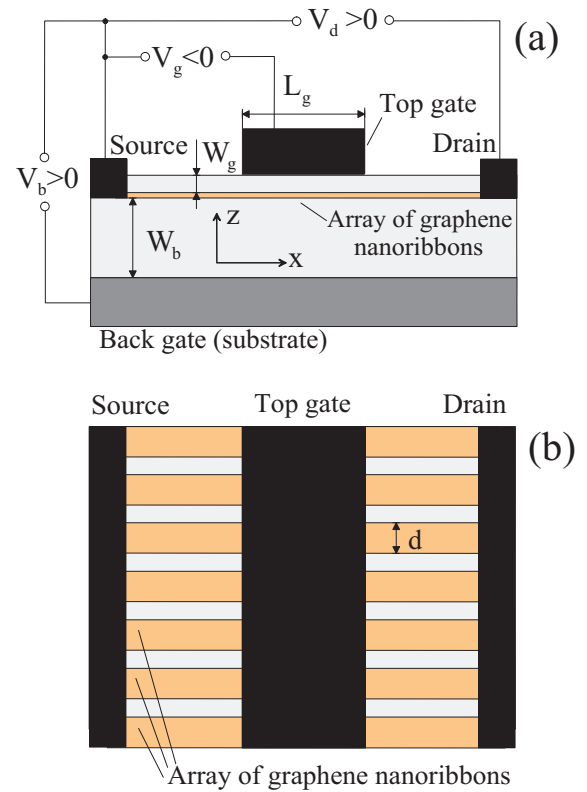}}
\caption{Schematic side (a) and top (b) views  of a GNR-FET structure.
}
\end{figure}


In this paper, we present an analytical device model for  a
graphene-nanoribbon FET (GNR-FET) and obtain the device characteristics. 
The GNR-FET under consideration
is based on a patterned graphene layer
which constitutes a dense  array of parallel
nanoribbons of width $d$ with the spacing
between the nanoribbons $d_s \ll d$. 
The nanoribbon edges are connected to the conducting pads
serving as the transistor source and drain.  
A highly conducting substrate plays the role of the back gate,
whereas 
the top gate serves to control the source-drain current.
The device structure is schematically 
 shown in Fig.~1.
Using the developed model, we calculate
the potential distributions
in 
the GNR-FET 
as a function of the back gate, top gate, and drain voltages,
$V_b$, $V_g$, and $V_d$ (reckoned from  the potential
of the source contact), respectively,
and the GNR-FET
dc  characteristics.
This corresponds to a GNR-FET in the common-source circuit.
The case of common drain will briefly be discussed as well.
For the sake of definiteness, the back gate voltage $V_b$ and
the top gate voltage $V_{g}$
are assumed to be, respectively,  positive and negative ($V_b > 0$
and $V_{g} < 0$) with respect to the potential
of the source contact, so we consider GNR-FETs with the channel of  n-type.

The paper is organized as follows. In Sec.~II,
the GNR-FET model is considered and the main equation
governing the spatial distributions of the electric potential
in the active region of the transistor channel is  given.
This equation is a consequence of the two-dimensional Poisson equation
in the weak nonlocality approximation~\cite{10,12,13}.
Section~III deals with the derivation of the relations between
the electron and hole densities in the channel and the
electric potential and the consequent calculations of
the potential spatial distributions depending on the drain and gate
(back gate and top gate) voltages. In Sec.~IV, using the obtained formulas for
the electric potential and the height of the barrier for electrons
at the minimum of the  potential, we obtain analytical expressions
for the source-drain current as a function of the drain and gate
voltages. In Sec.~V, we discuss the specific
of  GNR-FET operation 
in the circuits with common drain, the possible  role
of holes, and the limitations of the model used.
In Sec.~VI, we draw the main results.  

\section{GNR-FET model}

Graphene strips (nanoribbons) exhibit the energy spectrum
with a gap between the valence and conduction bands
depending on the nanoribbon width $d$:
\begin{equation}\label{eq1}
\varepsilon_{p,n}^{\mp} = \pm v\,\sqrt{p^2 + (\pi \hbar/d)^2n^2}.
\end{equation} 
Here $v \simeq 10^8$~cm/s is the characteristic velocity of the electron
(upper sign) and hole (lower sign)
spectra, $p$ is the momentum in along the nanoribbon, $\hbar$ is the reduced
Planck constant, and $n = 1,2,3,...$ is the subband index.
The quantization  corresponding to Eq.~(1)
of the electron  and hole 
energy spectra
in nanoribbons due to the electron and hole confinement
in one of the lateral directions results in 
the appearance of the band gap between the valence and conduction bands
and in 
a specific
 the density of states (DOS) as a function of the energy.

The electron density in different parts of the channel and,
therefore, the degree of the electron gas degeneration
essentially depend on the back gate voltage.
In contrast to graphene with zero energy gap,
the electron (hole) gas in  relatively narrow nanoribbons with
quantized energy spectrum and the pertinent energy gap
between the subbands in the valence and conduction band,
can become degenerate at fairly high back gate voltages.
In the following, we restrict ourselves by the consideration
of the GNR-FET operation under the condition
that the electron gas in the channel is nondegenerate.
Such a consideration is valid in the voltage range which
is sufficiently  wide in terms of practical applications. 
Thus, the back gate voltage is assumed to be 
not excessively high,
so that the electron density is moderate,
the electron gas in the  channel is nondegenerate, and
the electrons occupy only the lowest ($n = 1$) subband
in the conduction band nanoribbons. 
Notwithstanding this, it is suggested that
 the channel sections adjacent to
the source and drain contacts are highly conducting,
so that these section are equipotential.
The potentials are equal to the potentials
of the source and drain contacts,  i.e., equal
to $\varphi = 0$  and $\varphi = V_d$, respectively.

We shall consider the GNR-FET region under the top
gate (the device active region)
defined as follows: $- L_g \leq x \leq L_g/2$,
$-W_b \leq z \leq W_g$, where $L_g$ is the length of the top gate,
and 
$W_b$ and $W_g$ are the thicknesses of the layers between
graphene and the back and top gates, respectively.
Here, the axis $x$ is directed along the nanoribbons,
whereas the axis $z$ is directed perpendicular to the nanoribbons
and gate planes.

To find the potential distribution along the channel,
we use the following equation:
\begin{equation}\label{eq2}
\frac{(W_b + W_g)}{3}\frac{d^2\,\varphi}{d\,x^2}
- \frac{\varphi - V_b}{W_b} - \frac{\varphi - V_g}{W_g}
= \frac{4\pi e}{\ae}(\Sigma_{-} - \Sigma_{+})
\end{equation}
with the boundary conditions
\begin{equation}\label{eq3}
\varphi|_{x = -L_g/2} = 0, \qquad \varphi|_{x = L_g/2} = V_d.
\end{equation}
Here, $\Sigma_{-}$ and  
$\Sigma_{+}$ are the electron and hole sheet densities
in the channel, $e$ is the electron charge, and $\ae$ is the 
dielectric constant of the material separating the channel from the gates. 
Equation~(2), which  governs the electric potential, 
$\varphi = \varphi(x) = \psi(x,0)$,
in the channel is a consequence of the two-dimensional
 Poisson equation for the electric potential $\psi = \psi(x,z)$
for the active region under consideration
in the  weak nonlocality approximation~\cite{12,13}.
This equation provides solutions, which can be obtained from
the exact solution of the two-dimensional Poisson by the expansion
in powers of the parameter 
$\delta = [(W_b^3 + W_g^3)/45 (W_b + W_g){\cal L}^2] $, 
which is much smaller than unity in the situation under consideration.
Here ${\cal l}$ is the 
characteristic scale of the  lateral inhomogeneities (see Sec.~V).
The lowest approximation  in such an expansion corresponds
to the so-called gradual channel approximation proposed by W.~Shockley,
in which the first
term in the left-hand side of Eq.~(2) is neglected~\cite{14,15},
so that the relationship between the potential
in the channel and the electron and hole charge becomes
local.
Thus, Eq.~(2) can be used when $\delta < 1$ and, hence, when 
${\cal L} \gtrsim max\,\{W_b, W_g\}$.
The harnessing of the approximation under consideration
makes, in particular,  possible to study essentially nonuniform
potential distributions in the GNR-FET channel and
the short-gate
effects analytically. 
The weak nonlocality  approximation was effectively
used
previously in some previous papers (see, for instance, 
Refs.~\cite{10,12,13,16}).

Since the spacing, $d_s$ between the nanoribbons
is small, we disregard a small scale nonuniformity
of the electric potential distribution in the in-plane 
direction $y $.

\section{Potential distributions}

The application of negative top gate voltage
leads to the formation of a potential barrier in the channel under
this gate. This barrier determined by the gate voltage 
controls the source-drain current and, hence, is responsible of
the device operation as a transistor.

\subsection{Electron and hole densities}

Considering that the electron and hole
 gases are nondegenerate, and, hence, 
the electron and hole distribution functions 
in the  subbands with $n = 1$  near the source and drain
is given by
$$
f_p^{\mp} \simeq
\exp\biggl( \frac{\pm e\varphi \pm \varepsilon_F - \sqrt{v^2p^2 + \Delta^2/4}}{k_BT}\biggr),
$$
where $\varepsilon_F$ is the Fermi energy reckoned from
the middle of the energy band gap and
$\Delta = \varepsilon^{+}_{0,1} - \varepsilon^{-}_{0,1}
= 2\pi\,v\hbar/d$ is the value of the energy band gap, the electric potential, 
the electron densities and the Fermi energies
in the source and drain regions (marked by 
superscripts ``$s$'' and ``$d$'', respectively) are related to each other as  

$$
\Sigma_{\mp}^s = \frac{4k_BT}{\pi\hbar\,d\,v}
\exp\biggl(\pm\frac{\varepsilon_F^s + e\varphi}{k_BT}\biggr)
\int_{\xi_m}^{\infty}
\frac{d\xi\,\xi\exp(- \xi)}{\sqrt{\xi^2 - \xi_m^2}}
$$
\begin{equation}\label{eq4}
=\frac{2\Delta}{\pi\hbar\,d\,v}
\exp\biggl(\pm\frac{\varepsilon_F^s + e\varphi}{k_BT}\biggr)
 K_1\biggl(\frac{\Delta}{2k_BT}\biggr),
\end{equation}

$$
\Sigma_{\mp}^d = \frac{4k_BT}{\pi\hbar\,d\,v}
\exp\biggl[\pm\frac{\varepsilon_F^d + e(\varphi - V_d)}{k_BT}\biggr]
\int_{\xi_m}^{\infty}
\frac{d\xi\,\xi\exp(- \xi)}{\sqrt{\xi^2 - \xi_m^2}}
$$
\begin{equation}\label{eq5}
=\frac{2\Delta}{\pi\hbar\,d\,v}
\exp\biggl[\pm\frac{\varepsilon_F^d + e(\varphi - V_d)}{k_BT}\biggr]
 K_1\biggl(\frac{\Delta}{2k_BT}\biggr).
\end{equation}
Here, $\xi_m = \Delta/2k_BT$ and $K_1(\xi)$ 
is the modified Bessel function. 
Taking into account
the asymptotic behavior of the  Bessel function 
at $\xi \gg 1$, namely, $K_1(\xi) \simeq (\pi/2\xi)^{1/2}\exp(- \xi)$, 
for $\Delta \gg 2k_BT$, we obtain
\begin{equation}\label{eq6}
\begin{array}{l}
\Sigma_{\mp}^{s}
\simeq \displaystyle\frac{2\sqrt{\Delta\,k_BT}}{\sqrt{\pi}\hbar\,d\,v}
\exp\biggl(\pm\frac{\varepsilon_F^{s} + e\varphi}{k_BT}
- \frac{\Delta}{2k_BT}\biggr),\\
\Sigma_{\mp}^{d}
\simeq \displaystyle\frac{2\sqrt{\Delta\,k_BT}}{\sqrt{\pi}\hbar\,d\,v}
\exp\biggl[\pm\frac{\varepsilon_F^{d} + e(\varphi - V_D)}{k_BT}
- \frac{\Delta}{2k_BT}\biggr].
\end{array}
\end{equation}
Considering   that when $\varphi = 0$
and $\varphi = V_d$, $\Sigma_{-,0}^s  - \Sigma_{+,0}^s= \ae\,V_b/4\pi\,eW_b$
and  $\Sigma_{-,0}^d - \Sigma_{+,0}^d= 
\ae\,(V_b - V_d)/4\pi\,eW_b$, respectively, 
where the quantities with the index ``0'' are the electron
and hole densities in the immediate vicinity of the source and drain contacts,
we arrive at the following equation:
$$
\sinh\biggl(\frac{\varepsilon_F^{s}}{k_BT}\biggr) 
= \exp\biggl(\frac{\Delta}{2k_BT}\biggr)
$$
\begin{equation}\label{eq7}
\times\biggl[\frac{\ae}{16\sqrt{\pi}}\frac{eV_b}{\sqrt{\Delta\,k_BT}}
\biggl(\frac{d}{W_b}\biggr)\biggl(\frac{\hbar\,v}{e^2}\biggr)\biggr],
\end{equation}
$$
\sinh\biggl(\frac{\varepsilon_F^{d}}{k_BT}\biggr) 
= \exp\biggl(\frac{\Delta - 2eV_d}{2k_BT}\biggr)
$$
\begin{equation}\label{eq8}
\times\biggl[\frac{\ae}{16\sqrt{\pi}}\frac{e(V_b - V_d)}{\sqrt{\Delta\,k_BT}}\biggl(\frac{d}{W_b}\biggr)\biggl(\frac{\hbar\,v}{e^2}\biggr)\biggr].
\end{equation}
When $V_b = V_g = V_d = 0$, one obtains
$\varepsilon_F^{s} = \varepsilon_F^{d} = 0$,
so that the electron and hole densities are equal to their thermal 
value~\cite{17}
\begin{equation}\label{eq9}
\Sigma_T = \frac{2\sqrt{\Delta\,k_BT}}{\sqrt{\pi}\hbar\,d\,v}\,
\exp\biggl(- \frac{\Delta}{2k_BT}\biggr).
\end{equation}
For the effective operation of a GNR-FET with the $n$-type channel,
the electron densities in the sections of the channel
should be reasonably high (to provide necessary conductivity
of these sections). This requires the application of a sufficiently
high back gate voltage. 
If 
\begin{equation}\label{eq10}
\exp\biggl(-\frac{\Delta}{2k_BT}\biggr),\,\, 
 \ll \frac{V_b}{V_F},\,\, 
\frac{(V_b - V_d)}{V_F} < 1,
\end{equation}
where
\begin{equation}\label{eq11}
V_F = \frac{\sqrt{\Delta\,k_BT}}{e}
\biggl[\frac{8\sqrt{\pi}}{\ae}\biggl(\frac{W_b}{d}\biggr)
\biggl(\frac{e^2}{\hbar\,v}\biggr)\biggr],
\end{equation}
as follows from Eqs.~(7) and (8), $\varepsilon_F^s, \varepsilon_F^d
\gg k_BT$, and
\begin{equation}\label{eq12}
\begin{array}{l}
\varepsilon_F^{s} \simeq  \displaystyle\frac{\Delta}{2} + k_BT\,\ln\biggl(\frac{V_b}{V_F}\biggr),\\
 \varepsilon_F^{d} \simeq \displaystyle\frac{\Delta}{2} +  
k_BT\,\ln\biggl(\frac{V_b - V_d}{V_F}
\biggr).
\end{array}
\end{equation}
At $V_b$ and $(V_b - V_d)$ satisfying inequality~(10),
the electron gas is nondegenerate and  
the electron density markedly exceeds its thermal value as well as
the density of holes. 
In the case under consideration, 
the electron density in the pertinent sections of
the channel, as follows from Eqs.~(6) and (12), are given by

\begin{equation}\label{eq13}
\begin{array}{l}
\Sigma_{-}^{s}
\simeq \displaystyle\frac{\ae\,V_b}{4\pi\,e W_b}
\exp\biggl(\frac{ e\varphi}{k_BT}\biggr),\\ 
\Sigma_{-}^{d}
\simeq \displaystyle\frac{\ae\,(V_b - V_d)}{4\pi\,e W_b}
\exp\biggl[\frac{ e(\varphi - V_d)}{k_BT}\biggr].
\end{array}
\end{equation}

%
%

Since at $V_b > 0$, the electron density markedly
exceeds the hole density, we can neglect $\Sigma_h$
in the right-hand side of Eq.~(2).
The case when the hole gas under the top gate
becomes essential (at very high top gate voltages)
will be discussed in the following (in Section~V).
%
Considering the relationships between the electron density
and the electric potential, from Eq.~(2) one can arrive at
the following equations governing the potential distribution
in
the active region: 

$$
\frac{d^2\,\varphi}{d\,x^2}
-
\frac{3}{W_bW_g}\varphi = - \frac{3(V_b/W_b + V_g/W_g)}{(W_b + W_g)} 
$$
\begin{equation}\label{eq14}
+ \frac{3}{(W_b + W_g)}\frac{V_b}{W_b}
\exp\biggl(\frac{e\varphi}{k_BT}\biggr),
\end{equation}
$$
\frac{d^2\,\varphi}{d\,x^2}
-
\frac{3}{W_bW_g}\varphi = - \frac{3(V_b/W_b + V_g/W_g)}{(W_b + W_g)} 
$$
\begin{equation}\label{eq15}
+ \frac{3}{(W_b + W_g)}\frac{(V_b - V_d)}{W_b}
\exp\biggl[\frac{e(\varphi -V_d)}{k_BT}\biggr].
\end{equation}
The exponential dependences in the right-hand sides of Eqs.~(14)
and (15) are valid provided that 
inequality~(10) is satisfied,   in particular, if
the electron gas is nondegenerate.
The threshold value of the back gate voltage, at which
the degeneration of the electron gas occurs, can be estimated
as $V_b \simeq V_F$.
\begin{figure}
\centerline{\includegraphics[width=75mm]{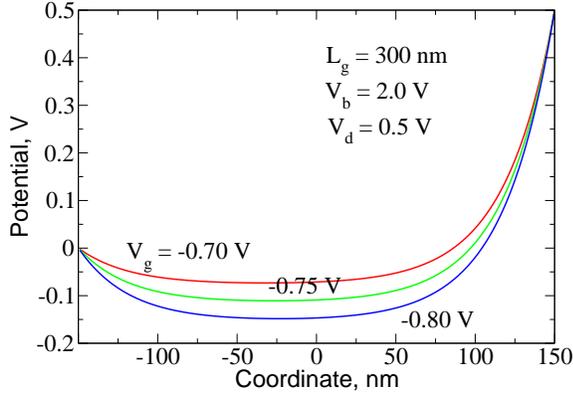}}
\caption{Spatial distributions of the potential 
at different top gate voltages.
}
\end{figure}

\begin{figure}
\centerline{\includegraphics[width=75mm]{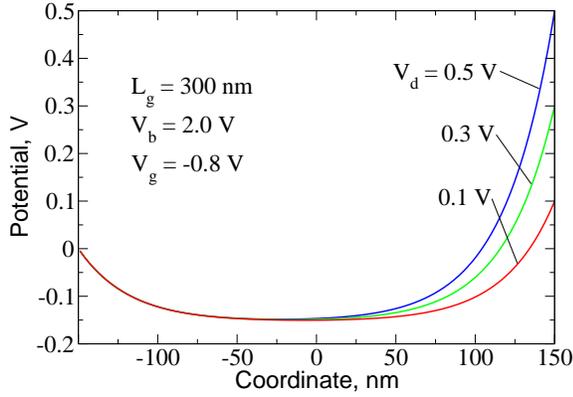}}
\caption{Spatial distributions of the potential 
at different drain voltages.
}
\end{figure}

\begin{figure}
\centerline{\includegraphics[width=75mm]{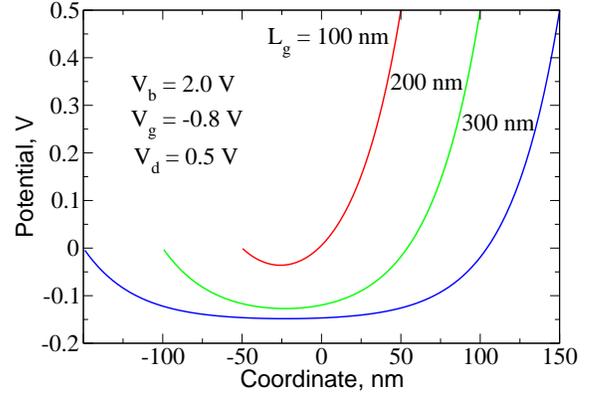}}
\caption{Spatial distributions of the potential 
in GNR-FETs with different gate lengths.
}
\end{figure}

\subsection{Potential distributions at low top-gate voltages}

When the top gate is negative ($V_g < 0$) and its
   absolute value $|V_g|$ is sufficiently small,
the modulus of the potential $|\varphi|$ can be not that large.
In this case, one can linearize Eqs.~(14) and (15) and present these equations in the following form:


$$
\frac{d^2\,\varphi}{d\,x^2} -
\frac{3}{W_bW_g}\biggl[1 + \frac{W_g}{(W_b + W_g)}\frac{eV_b}{k_BT}\biggr]
\varphi 
$$
\begin{equation}\label{eq16}
= - \frac{3V_g}{(W_b + W_g)W_g}, 
\end{equation}
$$
\frac{d^2\,\varphi}{d\,x^2} -
\frac{3}{W_bW_g}\biggl[1 + \frac{W_g}{(W_b + W_g)}\frac{e(V_b - V_d)}{k_BT}
\biggr]\varphi 
$$
\begin{equation}\label{eq17}
= - \frac{3(V_g/W_g - V_d/W_b)}{(W_b + W_g)}.
\end{equation}
%
%

Equations~(16) and (17) should be solved taking into account Eq.~(3)
and the matching of $\varphi$ and $d\varphi/d\,x$
at a certain point $x = x_m$
(such that $- L_g/2 < x_{m} < L_g/2$), at which $\varphi$ reaches a minimum.
Such a minimum definitely exist when $V_g < 0$.
At $V_d = 0$, Eqs.~(16) and (17) yield

\begin{equation}\label{eq18}
\varphi \simeq \frac{V_g\,W_b}{(W_b + W_g)}\,
\frac{\displaystyle\biggl[1 - \frac{\cosh(x/\lambda)}{\cosh(L_g/2\lambda)}\biggr]}
{\displaystyle\biggl[1 + \frac{W_g}{(W_b + W_g)}\frac{eV_b}{k_BT}\biggr]}.
\end{equation}
where
$$
\lambda = \Lambda \biggl/\sqrt{1 + \frac{W_g}{(W_b + W_g)}\frac{eV_b}{k_BT}}
\simeq \Lambda \biggl/\sqrt{\frac{W_g}{(W_b + W_g)}\frac{eV_b}{k_BT}}
$$
is the 
effective  screening length and $\Lambda = \sqrt{W_bW_g/3}$.
At $V_d = 0$, as follows from Eq.~(18), the function $\varphi$ exhibits
a minimum $\varphi = \varphi_{m0}$ at $x = 0$ with
$$
\varphi_{m0} \simeq \frac{V_g\,W_b}{(W_b + W_g)}\,
\frac{\displaystyle\biggl[1 - \frac{1}{\cosh(L_g/2\lambda)}\biggr]}
{\displaystyle\biggl[1 + \frac{W_g}{(W_b + W_g)}\frac{eV_b}{k_BT}\biggr]}
$$
\begin{equation}\label{eq19}
\simeq
\frac{V_g}{V_b}\frac{W_b}{W_g}\frac{k_BT}{e}.
\end{equation}
Here, we have taken into account that normally $V_b \gg k_BT/e$
and $L_g \gg \lambda$ (with $\lambda < \Lambda$).

At reasonable values of the  drain voltages $V_d$,
the variation of  $\varphi_m - \varphi_{m0}$ 
is equal to $V_d/2\cosh(L_g/2\lambda) \ll V_d$, i.e., is
very small due to $\cosh(L_g/2\lambda) \gg 1$
and can be disregarded, so that
\begin{equation}\label{eq20}
\varphi_m \simeq \varphi_{m0} 
\simeq \frac{V_g}{V_b}\frac{W_b}{W_g}\frac{k_BT}{e}.
\end{equation}
%
%
Comparing~$e|\varphi_m|$ given by Eq.~(20) with $k_BT$,
we find that Eqs.~(16) - (19) are valid
when
$|V_g|  \lesssim V_bW_g/W_b$

\subsection{Potential distributions at
high top-gate voltages}

At high back gate voltages, the quantity $\lambda$ playing the role of 
the screening length is rather small.
In this case,
the length of the regions   
 near the points
$x = \pm L_g/2$, in which the potential changes from $\varphi = 0$
to $|\varphi| > k_BT$ and from $\varphi = V_d$ to $|\varphi - V_d| > k_BT$,
is small in comparison with the top gate length $L_g$.
In such short regions near $x = \pm L_g/2$, the potential distribution
can still be describe by Eq.~(18).
However, in a significant part of the active region the electron charge,
which is in such a situation exponentially small, can be disregarded
and the last (exponential) terms in Eqs.~(14) and (15) can be omitted.
Taking this into account, at high top gate voltages, 
Eqs.~(14) and (15) in the central region
can be presented in the following form:
\begin{equation}\label{eq21}
\frac{d^2\,\varphi}{d\,x^2}
-
\frac{3}{W_bW_g}\varphi = - \frac{3( V_g/W_g + V_b/W_b)}{(W_b + W_g)}. 
\end{equation}
Solving Eqs.~(21) still using boundary conditions given by Eq.~(3), 
we arrive at  

%
$$
\varphi =  \biggl(V_g + V_b\frac{W_g}{W_b}\biggr)\frac{W_b}{(W_b + W_g)}\biggl[1 
- \frac{\cosh(x/\Lambda)}{\cosh(L_g/2\Lambda)}\biggr]
$$
\begin{equation}\label{eq22}
+ V_d\frac{\sinh[(x + L_g/2)/\Lambda]}{\sinh(L_g/\Lambda)}, 
\end{equation}
%


At $V_d = 0$, $\varphi$ exhibits a minimum at $x = 0$ and

\begin{equation}\label{eq23}
\varphi_{m0} \simeq
\biggl(V_g + V_b\frac{W_g}{W_b}\biggr)
\frac{W_b}{(W_b + W_g)}\,
\biggl[1 - \frac{1}{\cosh(L_g/2\Lambda)}\biggr].
\end{equation}
In the limit  $L_g \gg \Lambda$, the potential in the center of the channel
given by Eq.~(22) coincides with that obtained in the gradual channel
approximation. However, in real GNR-FET structures, $L_g$
can be comparable with $\Lambda$. In this case, the terms dependent
on $L_g$ can be important (in contrast to the case of low gate voltages
considered in the previous subsection).
At reasonable values of the drain voltage $V_d$ (sufficiently small
compared to $V_b$),
Eq.~(22) yields 
$$
\varphi_{m} \simeq
\biggl(V_g + V_b\frac{W_g}{W_b}\biggr)
\frac{W_b}{(W_b + W_g)}\,
\biggl[1 - \frac{1}{\cosh(L_g/2\Lambda)}\biggr]
$$
\begin{equation}\label{eq24}
+\frac{V_d}{2\cosh(L_g/2\Lambda)}. 
\end{equation}

Figures~2 and 3 show examples of the spatial distributions  (along the channel,
i.e.,
in the $x$-direction) of 
the electric potential in the active
region (under the top gate) calculated  
for a GNR-FET with $W_b = 100$~nm, $W_g = 30$~nm, and $L_g = 300$~nm at the
back gate voltage $V_b = 2.0$~V,
assuming different values of the top gate voltage $V_g$ and the drain
voltage $V_d$. As seen from Fig.~2, the potential distribution
markedly sags and the height of the barrier for electrons near
the center increases with increasing absolute value of the top gate
potential $|V_g|$. Figure~3 demonstrates, in particular, that in
the GNR-FET with chosen parameters the minimum value of the potential
$\varphi_m$ and, hence, the height of the barrier $-e\varphi_m$ 
for the electrons 
propagating from the source are virtually insensitive to the drain voltage.
In contrast, the potential barrier $e(V_d - \varphi_m)$  
for the electrons propagating from
the drain increases linearly with increasing $V_d$.
Figure~4 shows
the spatial distributions of the electric potential calculated for
the GNR-FETs with different gate lengths $L_g$.


The approach used in this subsection is valid
when $e|\varphi_m| \gg k_BT$, i.e., 
when $|V_g| - V_bW_g/W_b \gg k_BT(W_b + W_g)/eW_b \sim  k_BT/e$.

In the limit $L_g \gg \Lambda > \lambda$
the equations for $\varphi_m$ obtained above (in particular Eq.~(24))
coincide with those
obtained using the gradual channel approximation.  

\section{Calculation and analysis of current-voltage characteristics}

Considering that the source-drain current
is determined by the electrons
 overcoming the potential
barrier under the top gate, one can use the following
formula for the density of this current (per unit length):
 
\begin{equation}\label{eq25}
J = \frac{2e}{\pi\,\hbar\,d}\biggl(\int_{p_m^s}^{\infty}dp\,v_p f_p^s
- \int_{p_m^p}^{\infty}dp\,v_p f_p^d\biggr).
\end{equation}
Here,
$$
v_p = \frac{d\varepsilon_{p,0}^{+}}{d\,p}
 = v^2\frac{p}{\sqrt{v^2p^2 + \Delta^2/4}}
$$
is the velocity of the  electron   with
 momentum $p$
in the lowest subband of the nanoribbon conduction band
and $p_{m}^s$ and $p_{m}^d$ are the momenta of the electrons
with the energies $e|\varphi_m|$ and $e(|\varphi_m| + V_d)$, respectively.
We have taken into account that the nanoribbon array is dense: $d_s \ll d$.
Integrating in Eq.~(26), we arrive at

$$
J =
 v \biggl(\frac{\ae}{2\pi^{3/2}\, W_b}\biggr) 
\sqrt{\frac{k_BT}{\Delta}}\exp\biggl(\frac{e\varphi_{m}}{k_BT}\biggr)
$$
\begin{equation}\label{eq26}
\times\biggl[V_b - (V_b - V_d)\exp\biggl(- \frac{eV_d}{k_BT}\biggr)\biggr] 
\end{equation}
with $\varphi_m$ given by Eq.~(20) at  low top gate voltages
and by Eq.~(24) at moderate and high voltages, respectively.
Equation~(26)
corresponds to the average thermal electron velocity
$v_T = v \sqrt{4k_BT/\pi\,\Delta}$.
One can see that the thermal electron velocity 
as well as the pre-exponential factor in the formula
for the current depend on the band gap.
This is due to the dependence on the gap of the electron dispersion
and the electron velocity.
\begin{figure}
\centerline{\includegraphics[width=75mm]{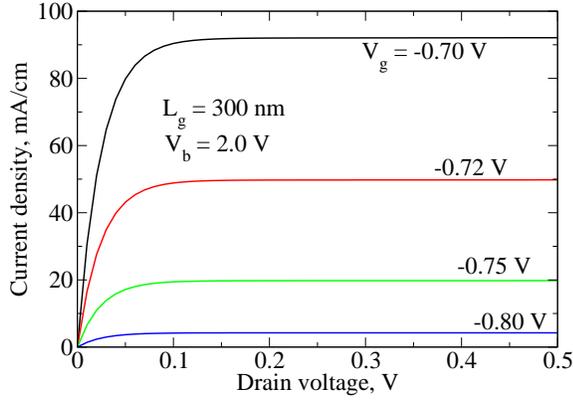}}
\caption{The source-drain current density versus drain
voltage  dependencies 
at fixed back gate voltage ($V_b = 2.0$~V)
and different top gate voltages $V_g$. 
}
\end{figure}
\begin{figure}
\centerline{\includegraphics[width=75mm]{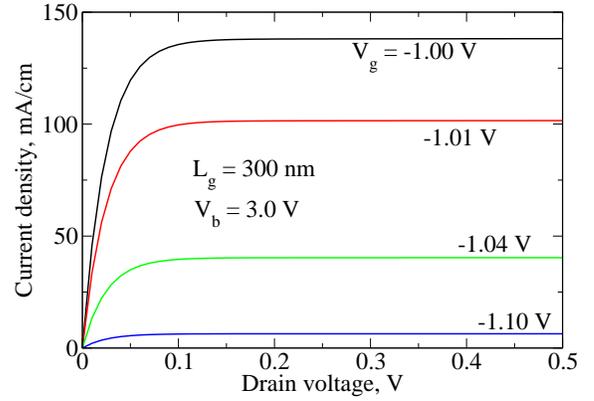}}
\caption{The same as in Fig.~5 but for $V_b = 3.0$~V.
}
\end{figure}

\subsection{Current versus drain voltage}

The dependence of the source-drain current on the drain voltage
is associated with 
the dependence of $\varphi_m$ on this voltage given
by Eqs.~(20) and (24)
and the voltage dependence of the last factor in the right-hand side
of Eq.~(26).
At low top gate voltages ($|V_g| \lesssim V_bW_g/W_b$ ), 
using Eqs.~(20) and (26), we obtain

\begin{equation}\label{eq27}
J \propto 
\biggl[V_b  - (V_b - V_d)\exp\biggl(- \frac{eV_d}{k_BT}\biggr)\biggr],
\end{equation}
i.e.,
\begin{equation}\label{eq28}
J \propto  V_d
\end{equation}
if $V_d \lesssim k_BT/e$ , and
\begin{equation}\label{eq29}
J  = const
\end{equation}
if  $k_BT/e \ll V_d \lesssim V_bW_g/W_b$. 

At moderate and high top gate voltages when $\varphi_m$ is given by Eq.~(24),
we arrive at the following dependence

$$
J \propto 
 \exp\biggl[\frac{eV_d}{2k_BT\cosh(L_g/2\Lambda)}\biggr]
$$
\begin{equation}\label{eq30}
\times\biggl[V_b - (V_b - V_d)\exp\biggl(- \frac{eV_d}{k_BT}\biggr)\biggr].
\end{equation}
At $V_d \lesssim k_BT/e$, Eq.~(30) yields the same linear
dependence on the drain voltage as that
described by  Eq.~(28). When $V_d \gg k_BT/e$,
one obtains 
\begin{equation}\label{eq31}
J \propto 
\exp\biggl[\frac{eV_d}{2k_BT\cosh(L_g/2\Lambda)}\biggr].
\end{equation}
The latter dependence resembles the dependence given by Eq.~(29),
but with the substitution of $\lambda$ by $\Lambda$. 
Since $\lambda \ll \Lambda$,
the source-drain current at moderate and high top gate voltages
is a less steeper function of the drain voltage than that
at low top gate voltages (compare Eqs.~(29) and (31)).

The  $J$ versus $V_d$ dependences 
for a GNR-FET with $L_g = 300$~nm
at the back gate voltages $V_g = 2$~V and $V_g = 3$~V
calculated for the different top gate voltages $V_g$
are shown in Figs~5 and 6.
Here, as in the previous and consequent figures,
we assumed that $\Delta = 0.4$~eV, $\ae = 4$,  $W_b = 100$~nm, 
 $W_g = 30$~nm, and $T = 300$~K.
Since the gate voltages were set to be relatively high ($V_b, |V_g| \gg k_BT/e
\simeq 0.025$V), Eqs.~(24) and (26) were used for the calculations.

As seen from Figs.~5 and 6, the source-drain current 
as a function of the drain voltage in a GNR-FET with relatively
long gate ($L_g = 300$~nm) exhibits saturation starting rather low
drain voltages. This is a consequence of very weak sensitivity
of the potential barrier for the electrons propagating from
the source to the drain voltage.

\subsection{Current versus gate voltages}

As follows from Eqs.~(20), (24), and (26),
the source-drain current as a function of the top gate voltage
is described by the following relations:

\begin{equation}\label{eq32}
J \propto 
 \exp\biggl(
\frac{V_g}{V_b}\frac{W_b}{W_g}\biggr)
\end{equation} 
at $|V_g| \lesssim V_bW_g/W_b$, and 

\begin{equation}\label{eq33}
J \propto 
 \exp\biggl[
\frac{eV_g}{k_BT}\frac{W_b}{(W_b + W_g)}
\biggl(1 - \frac{1}{\cosh(L_g/2\Lambda)}\biggr)
\biggr]
\end{equation} 
at $|V_g| > V_bW_g/W_b$.

Figures~5 and 6 show the transformation of the $J$ vs $V_d$ dependences
with varying top gate voltage in the range of the latter $|V_g| > V_bW_g/W_b$. 
These figures confirm
a strong sensitivity of the source-drain current to
the gate voltages.
As follows from Eqs.~(32) and (33), 
the dependence of  the source-drain current on the top gate voltage
is much steeper in the range high top gate voltages $|V_g| \gtrsim V_bW_g/W_b$
than at  $|V_g| \lesssim V_bW_g/W_b$, i.e.,  when the central region of the channel becomes essentially depleted.
Figure~7 shows the $J$ versus $V_g$ characteristics
obtained using Eq.~(26) in which $\varphi_m$ was
determined  by Eq.~(20) (for small $|V_g|$)
and by Eq.~(24) (for large $|V_g|$), respectively.
\begin{figure}
\centerline{\includegraphics[width=75mm]{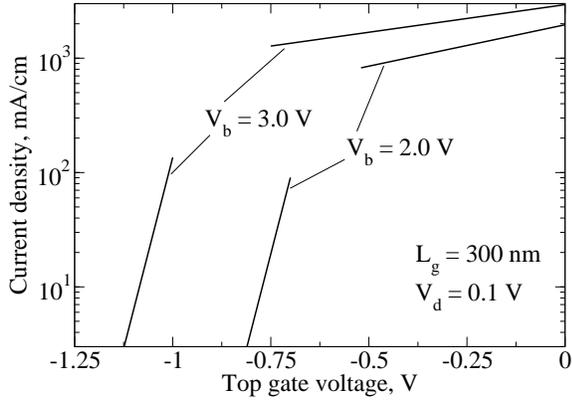}}
\caption{The source-drain current density as a function of the top gate voltage
at different back gate voltages. 
}
\end{figure}

The ratio  the transconductance $g = \partial J/\partial V_g$ to
the source-drain current $J$ at high top gate voltages $|V_g| > V_bW_g/W_b$
to the same ratio at low top gate voltage $|V_g| < V_bW_g/W_b$
is given by
\begin{equation}\label{eq34}
\biggl(\frac{g}{J}\biggr)_{high}
\biggr/\biggl(\frac{g}{J}\biggr)_{low}
\simeq  \frac{eV_b}{k_BT}\frac{W_g}{(W_b + W_g)} \gg 1.
\end{equation} 

The source-drain current versus the back gate dependence 
at high top gate and drain voltages
is given by
\begin{equation}\label{eq35}
J \propto 
 V_b \exp\biggl[
\frac{eV_b}{k_BT}\frac{W_g}{(W_b + W_g)}
\biggl(1 - \frac{1}{\cosh(L_g/2\Lambda)}\biggr)
\biggr].
\end{equation} 
The source-drain current increases
with increasing back gate voltage
due to the pertinent
 increase in the electron density in all regions of the channel.


\subsection{Short-gate effects}

As follows from the above formulas,
the current-voltage characteristics 
exhibits pronounced dependence on the top gate length.
This can be attributed to an essential dependence
of the height of the potential barrier for the electrons 
propagating from the source  on the top gate length (as, in particular,
 seen from Fig.~4)
and on the drain voltage. 
As a result, the  $J$ versus $V_d$ 
dependence (see Eqs.~(29) and (31)) becomes markedly steeper
 when
the top gate length decreases, particularly, when the latter
 becomes comparable with $\Lambda$ (or $\lambda$).

In particular, as seen from Eqs.~(29) and (31),
in a GNR-FET with
a long top gate,
 the source-drain current saturates when $V_d$ becomes
larger that $k_BT/e$, whereas
in a GNR-FET with $L_g$ comparable with $\Lambda$,
 the source-drain current markedly increases with
increasing $V_d$ even at rather large values of the latter.
This is confirmed by the current-voltage characteristics
calculated  for GNR-FETs with a ``long'' ($L_g = 300$~nm)
and a ``short''($L_g = 100$~nm) top gates shown in
Fig.~7. 
\begin{figure}
\centerline{\includegraphics[width=75mm]{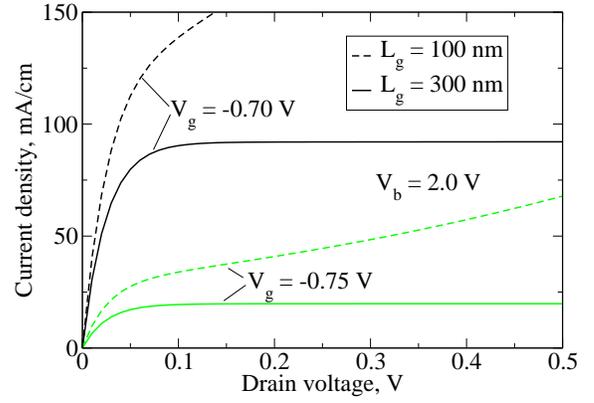}}
\caption{The source-drain current density  as a function of the drain voltage
in GNR-FETs with different gate lengths at 
different
top gate voltages: solid lines - $L_g = 300$~nm ("long" gate) and
dashed lines - $L_g = 100$~nm ("short" gate). 
}
\end{figure}

\section{Discussion}
\subsection{GNR-FET in  common-drain circuit}

Above we considered the case of GNR-FETs in the common-source
circuit (see Fig.~1), so that
the voltage  between the back gate 
and the source and between the top gate and the sourse are equal to $V_b$
and $V_g$, respectively.
In the case of GNR-FETs in the common-drain circuit,
the voltages $V_b$ and $V_g$ are  applied between that back gate and the drain
and between the top gate and drain.
In such a situation,
in the above formulas one needs to replace
$V_b$ by $V_b + V_d$ and   $V_g$ by $V_g + V_d$. 
In this case, Eqs.~(24) and (26) should be replaced by the following equations: 
$$
\varphi_{m} \simeq
\biggl[\frac{(V_g + V_bW_g/W_b)W_b}{(W_b + W_g)} + V_d\biggr]
\biggl[1 - \frac{1}{\cosh(L_g/2\Lambda)}\biggr]
$$
\begin{equation}\label{eq36}
+\frac{V_d}{2\cosh(L_g/2\Lambda)} 
\end{equation}
and
$$
J =
 v \biggl(\frac{\ae}{2\pi^{3/2}\, W_b}\biggr) 
\sqrt{\frac{k_BT}{\Delta}}\exp\biggl(\frac{e\varphi_{m}}{k_BT}\biggr)
$$
\begin{equation}\label{eq37}
\times\biggl[V_b + V_d -  V_b\exp\biggl(- \frac{eV_d}{k_BT}\biggr)\biggr]. 
\end{equation}
%
The main difference between the GNR-FET current-voltage
characteristics in the common source circuit
and in the common drain circuit
is that in the latter case
the source-drain current is a rather steep function
of the drain voltage even at $V_d \gg k_BT/e$.
Indeed, as seen from Eq.~(31),
in the GNR-FETs with $L_g \gg \Lambda$,
the current tends to saturation in the voltage range
$V_d \gg k_BT/e$. However, Eqs.~(36) and (37)
valid in the case of the common drain circuit,
give rise to an exponential voltage dependence, which at 
$V_d \gg k_BT/e$ becomes as follows:
\begin{equation}\label{eq38}
J \propto
\exp\biggl(\frac{eV_d}{k_BT}\biggr).
\end{equation}

\subsection{Role of holes}

At elevated top gate and drain voltages, the top of the valence band
in the central section of the channel
can become higher that the Fermi levels in the side sections.
In this case, a significant amount of holes can occupy
the central section of the channel, 
so that a $p$-region can be  formed
in this section.
Such an effect can occur when
$e|\varphi_m| > \Delta/2 + \varepsilon_F^s$
or $e|\varphi_m| + eV_d > \Delta/2 + \varepsilon_F^d$.
Since in the case of nondegenerate electron gas in the channel 
$\varepsilon_F^s, \, \varepsilon_F^d < \Delta/2$,
the condition, at which the hole gas under the top gate might be
essential, can be presented as 
$|\varphi_m|,\,  |\varphi_m| + V_d \gtrsim \Delta/e$.
%
As an example, assuming that $V_d \ll V_b$ and setting
$\Delta = 0.4$~eV, $W_g = 0.3 W_b$, and $V_b = 2 - 3$~V
one obtains that the hole effects can be essential if 
$|V_g| \gtrsim 1$~V.
It implies that the hole gas can affect the GNR-FET
characteristics at elevated but realistic values of the top gate voltages.
The  hole charge under the top gate results in slowing down
of the  decrease of the source-drain current with increasing $|V_g|$.
At  higher top gate and/or drain voltages, 
the tunneling between
the $n$-regions in the channel not covered by the top gate
and the $p$-region under this gate becomes tangible. As a result,
the source-drain current can increase dramatically.~\cite{6,7,8,9,10} 
The tunneling current between in the degenerate $n$-
and $p$-sections in the channel of a graphene FET   was studied 
previously.~\cite{6,10}
In the GNR-FETs with a marked energy band gap and
a nondegenerate electron gas in
the side sections of the channel ($n$- regions),
the tunneling current can become essential at fairly high
voltages. 
Indeed, taking into account
that the probability of the interband
tunneling between the subbands with $n = 1$
is given by~\cite{7,8} $t_t = exp(-\pi^3\hbar\,v/eE_md^2)$,
where $E_m$ is the characteristic electric field
in the n-p and p-n junctions arisen due to the population
of the central section of the channel by holes.
This field we estimate as 
$E_m = |\varphi_m|/\Lambda = |\varphi_m|\sqrt{3/W_bW_g}$
in the n-p jinction and
$E_m = (|\varphi_m| + V_d) \sqrt{3/W_bW_g}$ in the p-n junction.
Using the above estimates, one can find that
the tunneling might be essential at the values
of $|\varphi_m|$ and/or $V_d$ exceeding the characteristic
tunneling voltage $V_t = (\pi/4\sqrt{3})\Delta^2\sqrt{W_bW_g}/e\hbar\,v$.
Setting $W_b = 100$~nm, $W_g = 30$~nm, and $\Delta = 0.4$~eV,
we obtain $V_t \simeq 6.4$~V. If $V_d \ll V_b$ and $V_b = 2 - 3$~V,
this implies that the interband tunneling might become significant
at $|V_g| \gtrsim 9$~V.
However, the quantitative study of
the tunneling effects in GNR-FETs with narrower energy gap, in which
the tunneling can occur at moderate voltages,
 requires a separate
careful consideration
which is beyond the scope of this paper.

\subsection{Main limitations of the model}

The validity of the device model used above is limited by the following:\\
(1) Applicability of the weak nonlocality approximation
which requires that the parameter 
$\delta = [(W_b^3 + W_g^3)/45 (W_b + W_g)]{\cal L}^2 \ll 1$.
This problem might arise only in the case of  GNR-FETs with rather short
top gate length when the terms in Eqs.~(23) and (24)
containing $\cosh(L_g/2\Lambda)$
are essential.
In the most interesting case, as follows from Eq.~(22),
${\cal L} = \sqrt{W_bW_g/3}$. In such a case,
$\delta = [(W_b^3 + W_g^3)/15 (W_b + W_g)]W_bW_g \ll 1$.
In particular, the latter yields $\delta = 1/15 \simeq 0.067$ if $W_b = W_g$
and $\delta \simeq 0.175$ if $W_g/W_b = 0.3$.
Using the Poisson equation in the weak nonlocality approximation
and retaining the next term in the expansion of the two-dimensional
Poisson equation over $\delta$,
one can find that the value $L_g/2\Lambda)$
in Eq.~(24) should be replaced by fairly
close value $L_g\sqrt{(1 + \delta)}/2\Lambda \simeq L_g(1 + \delta/2)/2\Lambda$.\\
(2) Not excessively high back gate voltages ($V_b < V_F$). This is due
an assumption that the electron gas in the channel is nondegenerate. 
Setting $d = 3$~nm, $W_b = 100$~nm, $ae = 4$, $v = 10^8$~cm/s,
$\Delta = 0,4$~eV, and $T = 77 - 300$~K, we obtain
the following  estimate $V_F \simeq 14.0 - 27.6$~V.
This estimate, as well as the smallness of $\exp(- \Delta/k_BT)$,
imply
that inequality~(10) is satisfied
and, hence, Eqs.~(14) and (15) are valid in fairly wide
range of the back gate  voltages, which definitely covers
the range of voltages used in real devices.\\
(3) Not excessively high top gate and drain voltages,
so that the effects associated with holes under the top gate
can be disregarded (see the previous Subsection).\\
(4) The device structures with relatively week scattering of electrons
on impurities and phonons in the active region of the channel. The 
electron scattering in this region can give rise to a decrease in
the pre-exponentional factor in the formula for the source-drain current
in comparison to that in  Eq.~(26). However, the main voltage
dependences obtained above and those which can be found using
the drift-diffusion model of the electron transport in the active region
of the channel should be essentially the same. 
In some sense, the situation is similar to that which occurs in 
 the theory of the Schottky diodes and conventional p-n- junctions
 when the so-called
thermionic and diffusion models
are compared~\cite{14,15}.

One can see that the device model used above  adequately describes the
operation of GNR-FETs with realistic parameters at reasonable
applied voltages.
\section{Conclusions}

In conclusion, we developed an analytical device model for GNR-FETs.
The GNR-FET 
current-voltage characteristics, i.e., the dependencies of 
the source-drain current
versus the drain voltage as well as the back gate and top gate voltages,
 were calculated for the devices with different geometric parameters 
 (thicknesses of the gate layers and the top gate length)
  using this
model. In particular, we showed that shortening of the top (controlling)
gate dives rise to a substantial transformation of the 
GNR-FET current-voltage characteristics (the short-gate effect). 
The model can be used for the GNR-FET optimization.

\section{Acknowledgment}
The work was supported by CREST, the Japan Science and 
Technology Agency, Japan.


\end{document}